\pdfoutput=1
\PassOptionsToPackage{sort&compress}{natbib}
\documentclass[5p]{elsarticle}
\usepackage{amsmath,amssymb}
\usepackage{graphicx}
\usepackage{bm}
\usepackage{microtype}
\usepackage[hidelinks]{hyperref}

\journal{Materials Today Quantum}

\begin{document}

\begin{frontmatter}

\title{Terahertz detection of magnon chirality in polycrystalline CsCuCl$_3$}

\author{Juan S. Rodr\'{\i}guez-Hern\'{a}ndez\fnref{eq}}
\author{Gabriel F. Freitas\fnref{eq}}
\author{Nicolas M. Kawahala}
\author{Rafael S. Freitas}
\author{Felix G. G. Hernandez\corref{cor1}}
\ead{felixggh@if.usp.br}
\fntext[eq]{These authors contributed equally to this work.}
\cortext[cor1]{Corresponding author.}
\address{Instituto de F\'{\i}sica, Universidade de S\~{a}o Paulo, 05508-090,
S\~{a}o Paulo, SP, Brazil}

\begin{abstract}
In antiferromagnets, the handedness of magnon precession is usually measured on single crystals. In a pressed pellet of the chiral helimagnet CsCuCl$_3$, whose grains carry both structural handednesses and all orientations, we found that, from 5 to 7~T, the sub-terahertz magnon absorbs one circular polarization more than 20 times as strongly as the other, which rules out precession about each grain's own axis. Spin-wave theory attributes this to the Zeeman frequency, here two to four times the exchange-enhanced gap, which locks the precession axis to the field in every grain.  Magnon chirality in antiferromagnets is thus accessible without a single crystal.
\end{abstract}


\end{frontmatter}

\section{Introduction}\label{sec:intro}

A magnon carries angular momentum. In antiferromagnets and noncollinear magnets the modes come in pairs of opposite handedness \cite{Keffer1952,Rezende2019}, which fixes the circular polarization a mode absorbs, the sign of the spin current it pumps \cite{Vaidya2020} and the phonons it hybridizes with \cite{Cui2023}. This handedness, the sense of precession, is what we call magnon chirality; the term is also used for magnons with nonreciprocal dispersion \cite{Lin2026}. Chiral magnons have been excited and detected electrically \cite{Wang2024} and seen to propagate with a definite handedness \cite{Xu2026}; in altermagnets the splitting of opposite-handed branches is a predicted signature \cite{Smejkal2023} that has since been observed \cite{Liu2024,Biniskos2025}.

Measuring that handedness demands a polarization-resolved probe and, in practice, a single crystal. Polarized inelastic neutron scattering has resolved the polarization of the magnons of yttrium iron garnet \cite{Nambu2020}, circularly polarized resonant inelastic x-ray scattering resolves chiral phonons \cite{Ueda2023} and individual antiferromagnetic domains \cite{Biniskos2025}, and circularly polarized terahertz absorption uses oriented crystals \cite{Schilberth2026}. A mode's handedness is defined with respect to an axis, so averaging over the orientations of that axis is expected to blur it. Saturated polycrystalline ferrites are the familiar exception: the field aligns the magnetization of every grain, and with it the precession axis, so that only one circular polarization is resonant, the basis of ferrite Faraday rotators and isolators at microwave frequencies \cite{Polder1949,Hogan1952}. Ferrimagnetic films that are not single crystals behave alike: the handedness of their spin waves has been followed through the compensation of the sublattice angular momenta in Gd--Co alloys \cite{Kim2020} and switched and read electrically in sputtered Py/Gd multilayers \cite{Liu2022}, in both cases on a net magnetization that the field aligns. An antiferromagnet has no spontaneous magnetization for the field to align and, as a result, each grain precesses about axes set by its own lattice, and the field takes over only when the Zeeman frequency exceeds the gap, which exchange enhances.

Chiral magnets, the setting of skyrmion and soliton textures \cite{Muhlbauer2009,Togawa2012,Fert2017} and of nonreciprocal transport \cite{Aoki2019}, add a handedness of the structure and of the spin helix, conventionally measured by diffraction of circularly polarized x rays \cite{Tanaka2008,Lovesey2008,Usui2014,Lang2004}. In CsCuCl$_3$ both are fully characterized: below a cooperative Jahn--Teller transition at $\sim\!423$~K the lattice adopts one of the enantiomorphic space groups $P6_122$/$P6_522$ \cite{Schlueter1966,Hirotsu1975,Hirotsu1977}, circularly polarized x-ray diffraction has fixed the crystal handedness and imaged domains of both handednesses \cite{Kousaka2009,Ohsumi2013}, polarized neutrons have shown the spin helix to be monochiral in homochiral crystals \cite{Kousaka2017}, and resonant x-ray diffraction has resolved orbital and magnetic chiralities \cite{Ueda2022}. The handedness of its excitations has not been measured. Since the helix is generated by antisymmetric exchange \cite{Dzyaloshinsky1958,Moriya1960,Adachi1980}, the crystal handedness fixes
the spin handedness \cite{Kousaka2017,Marty2008}. Crystals grown from solution
contain domains of both handednesses \cite{Tanaka1985}, and thermal cycling through the Jahn--Teller transition nucleates both \cite{Hirotsu1977,Ueda2025}. A pellet pressed from crushed crystals thus contains both handednesses and all orientations of the crystal axes. We show that its magnon nonetheless absorbs only one circular polarization, and that the field, not the crystal axes, sets its precession axis.

\begin{figure}[tp]
\includegraphics[width=\columnwidth]{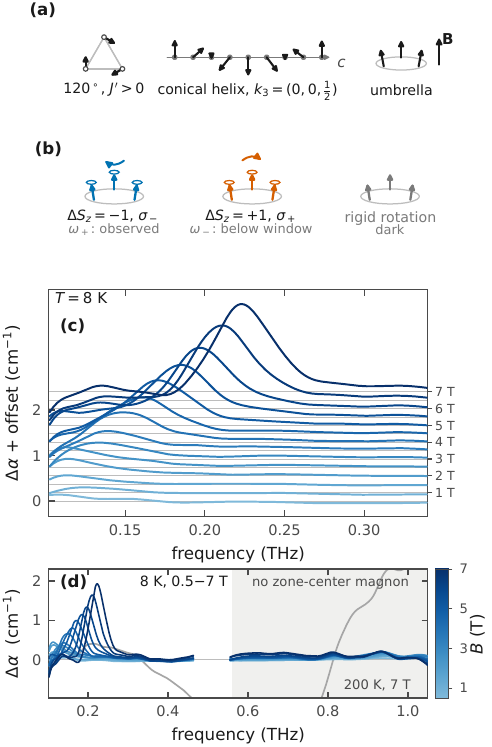}
\caption{\label{fig:one}
Structure, modes and field-induced absorption of CsCuCl$_3$.
(a)~$120^\circ$ order of the ferromagnetic chains (left), conical helix along
$c$ (center) and umbrella for $\bm{B}\parallel c$ (right).
(b)~Zone-center modes of the umbrella.
(c)~$\Delta\alpha$ [Eq.~(\ref{eq:dalpha})] at 8~K, 0.5--7~T in 0.5~T steps,
offset.
(d)~The same spectra up to 1.05~THz without offset; shaded, the range in which
Eq.~(\ref{eq:H}) has no zone-center magnon; grey, 200~K at 7~T (partly off
scale). The phonon edge at 0.47--0.56~THz is masked.}
\end{figure}

\section{Structure and modes}\label{sec:structure}

\begin{figure*}[tp]
\includegraphics[width=\textwidth]{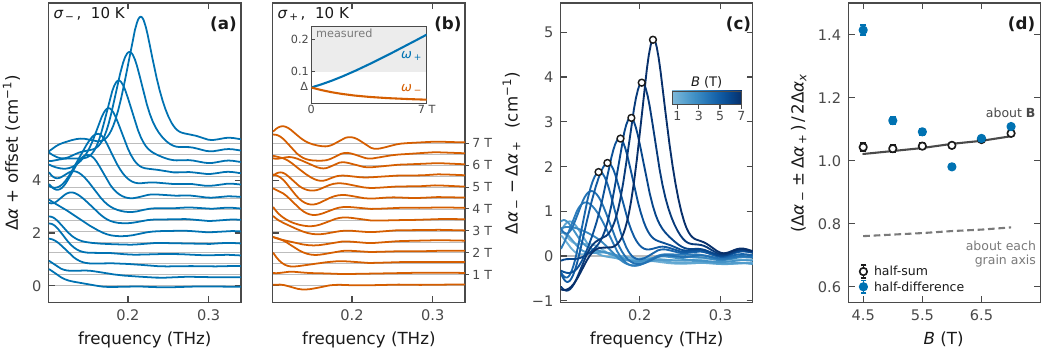}
\caption{\label{fig:two}
Circular channels at 10~K.
(a)~$\Delta\alpha_-$ and (b)~$\Delta\alpha_+$, 0.5--7~T in 0.5~T steps, with identical offsets and scales.
Inset of (b): illustration for the two branches of Eq.~(\ref{eq:branches}) at 10~K for $\bm{B}\parallel c$, with the single-crystal gap scaled to 10~K ($\Delta = 50$~GHz) and $g_c = 2.085$; shaded, the measured range, above 0.10~THz. $\omega_+$ ($\sigma_-$) enters it above 2.6~T, while $\omega_-$ ($\sigma_+$) $= \Delta^2/\omega_+ \le \Delta$ stays below it for any gap $\Delta_0 < 198$~GHz.
(c)~Dichroism spectrum $\Delta\alpha_- - \Delta\alpha_+$ at the same fields; circles indicate its value at the fitted line center.
(d)~Half-sum (open) and half-difference (filled) of the channels at the line center over the $x$-polarized absorption of the same runs, with the calculated values for precession about $\bm{B}$ (solid; both coincide) and about each grain's $c$ axis (dashed: half-difference; its half-sum is within 0.04 of the solid line).
}
\end{figure*}

In CsCuCl$_3$ the Cu$^{2+}$ spins ($S = 1/2$) form ferromagnetic chains along $c$, coupled antiferromagnetically by $J'$ on a triangular lattice [Fig.~\ref{fig:one}(a)]. Below $T_N = 10.7$~K they order in the $120^\circ$ state, which the Dzyaloshinskii--Moriya (DM) interaction twists by $5.1^\circ$ per Cu into a helix along $c$ \cite{Adachi1980}; a component along $c$ modulated at $\bm{k}_3 = (0,0,1/2)$ makes the helix conical \cite{Ueda2022}. A field along $c$ cants the three sublattices into an umbrella, whose three zone-center magnons differ in how the sublattices share the precession phase [Fig.~\ref{fig:one}(b)]. For relative phases $+120^\circ$ and $-120^\circ$ the transverse moments add up to a net moment rotating about $\bm{B}$ in opposite senses. These modes carry $\Delta S_z = \pm1$, absorb opposite circular polarizations and are split by the Zeeman frequency $\gamma B$. The rigid rotation of the umbrella has no net transverse moment and is dark, so the selection rule follows from the $120^\circ$ geometry. The mode observed here is the upper one, $\omega_+$ ($\Delta S_z = -1$), which absorbs $\sigma_-$; its partner $\omega_-$ lies below our window [inset of Fig.~\ref{fig:two}(b)].

\section{Experiment}\label{sec:experiment}

Crystals of CsCuCl$_3$ grown by slow evaporation of a stoichiometric CsCl--CuCl$_2$ solution in HCl \cite{RodriguezHernandez2024} were crushed and pressed into a pellet of thickness $L = 1.3$~mm (Supplementary Material, Section~S6.1). Transmission was measured by terahertz time-domain spectroscopy in the Faraday geometry \cite{Hernandez-review,m693-z3kk,dlh5-6ffv}, $\bm{k}\parallel\bm{B}$, over
0.10--1.70~THz, in fields to 7~T and at 1.5--300~K. The time traces were apodized with a Gaussian window of $\sigma_t = 10$~ps centered on the pulse \cite{Marulanda2025-wb}, which limits the resolution to 37.5~GHz and reduces the echo of the pellet, 21~ps behind the pulse, to 11\%. For the circular channels the incident radiation was linearly polarized along $x$; after the sample the field was measured along the two diagonals, and its $x$ and $y$ components were computed and transformed to the circular basis, so that both helicities $\sigma_\pm$ come from the same measurement \cite{Kawahala-npj2D}. The two analyzer settings were measured in separate runs, and each spectrum is referred to the zero-field spectrum of its own run. 
Each spectrum is referenced to zero field at the same temperature,
\begin{equation}
\Delta\alpha(\omega,B) = -\frac{1}{L}\,
\ln\!\left[\frac{T(\omega,B)}{T(\omega,0)}\right],
\label{eq:dalpha}
\end{equation}
which cancels the phonons \cite{Petzelt1981}, the Fabry--P\'{e}rot structure
and the instrument response; no fitted oscillator is subtracted, and a
zero-field dichroism would cancel too. Lines are fitted
with a Lorentzian on a quadratic baseline over 0.105--0.40~THz, with the spread
over baseline orders and windows as systematic error, and a line center is used
only where the whole line lies inside the experimental window. Calculated spectra
are multiplied into the measured traces, processed and fitted in the same
way, so that the fit bias, a downward shift of the fitted centers by about 4~GHz near
0.2~THz caused by the time window, cancels in every comparison with the calculation
(Supplementary Material, Section~S6.1).

\section{Results}\label{sec:results}

\subsection{Magnon line and selection rule}\label{sec:line+selection}

Below $T_N$ a single line appears in $\Delta\alpha$ and
moves up with field [Fig.~\ref{fig:one}(c)]. At 8~K it goes from $147 \pm 6$~GHz
at 4~T, the lowest field used, to
$224.6 \pm 0.5$~GHz at 7~T, where it peaks at 1.92~cm$^{-1}$, a 22\% dip in
transmitted intensity, with a width of 40~GHz, close to the resolution. Up to 1.05~THz, apart from a masked phonon
edge, it is the only resonance [Fig.~\ref{fig:one}(d)]. Above $\sim\!0.5$~THz
the model of Section~\ref{sec:model} has no zone-center magnon, so the field dependence there bounds
the non-magnetic response of the same run: over 0.56--1.05~THz at 7~T it
averages 0.12~cm$^{-1}$, peaks at 0.21~cm$^{-1}$ and carries no resonance
(Supplementary Material, Section~S6.3).

At 10~K and 7~T [Figs.~\ref{fig:two}(a) and \ref{fig:two}(b)] the magnon appears in $\Delta\alpha_-$ with a peak of 4.79~cm$^{-1}$, while $\Delta\alpha_+$ shows no line and carries $-0.05$~cm$^{-1}$ at the same frequency. The dichroism spectrum
$\Delta\alpha_- - \Delta\alpha_+$ peaks at the mode from 4.5~T on and follows
it from 0.15 to 0.22~THz, growing from 1.9 to 4.8~cm$^{-1}$ [Fig.~\ref{fig:two}(c)],
which neither a polarimetric offset nor acquisition drift can do.

\subsection{Precession axis}\label{sec:axis}

The size of the dichroism decides what sets the precession axis. Linear light is an equal mix of the two helicities, so a mode whose moment circles about $\bm{B}$ absorbs twice the linear value in one channel and nothing in the other, in every grain. A grain precessing about its own $c$ axis, tilted by $\theta$ from $\bm{k}$, would show the light an ellipse, and the helicities would absorb $(1 \pm |\cos\theta|)^2/2$ in units in which a circle about $\bm{k}$ gives 2 and 0; over random orientations the channel ratios become 1.75 and 0.25 (Supplementary Material, Section~S3). The circular runs absorb 11--26\% more than the linear sweeps interpolated to 10~K, so Fig.~\ref{fig:two}(d) refers the half-sum and half-difference of the channels at the line center to the $x$-polarized absorption of the same runs, and compares them with both predictions passed through our processing. The half-sum, 1.04--1.09, matches the predicted 1.02--1.08 within 0.02. The half-difference, 0.98--1.13 from 5 to 7~T, matches precession about $\bm{B}$, 1.03--1.08, and excludes precession about each grain's axis, 0.77--0.79. Equivalently, at the line center the dark channel carries between $-0.04$ and $+0.03$ of the active one, so the active channel is more than 20 times stronger, where grain axes would give 0.14, a factor of 7; only at 4.5~T, where the line sits at the lower edge of the window over a weak structure that the dark channel shows near 0.15~THz at most fields, does it reach $-0.15$. Fitting the dark channel instead with the line shape of the active one gives amplitudes between $-0.06$ and $+0.06$ of the active line averaged over 5.5--7~T, and between $-0.17$ and $+0.13$ at single fields; we take this spread, which comes from the same structure, as the systematic uncertainty of the dark fraction (Supplementary Material, Section~S6.3). The Faraday rotation accompanying the dichroism, which does not depend on the dark channel at the line, separates the lines seen through the two analyzers by 26--33~GHz at 6--7~T, as calculated for precession about $\bm{B}$ (28--31~GHz; grain axes, 23--24~GHz; Supplementary Material, Section~S6.3). The field, not the grain axis, sets the precession axis.

\begin{figure*}[tp]
\includegraphics[width=\textwidth]{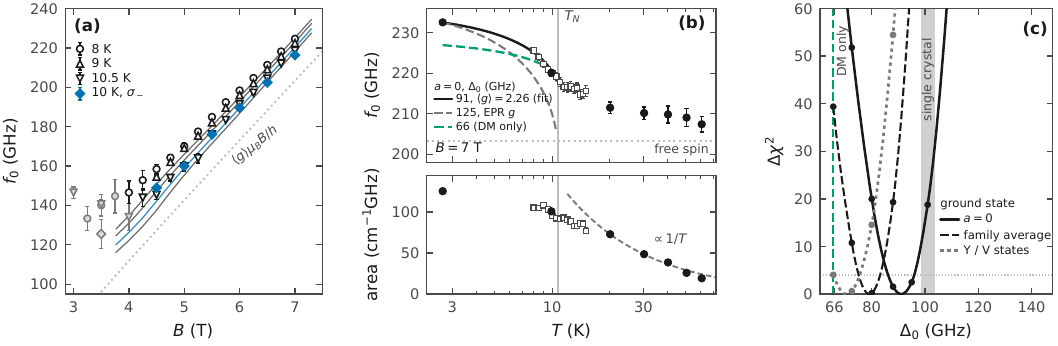}
\caption{\label{fig:three}
Line position and the gap.
(a)~Line center versus field for the four field sweeps (grey, not used), with
Eq.~(\ref{eq:H}) fitted to the 7~T line of (b) ($a = 0$, $\Delta_0 = 91$~GHz,
$\langle g\rangle = 2.26$) at the temperature of each sweep; dotted, free-spin line of the EPR $g$ tensor.
(b)~Center (top) and integrated intensity (bottom) of the 7~T line versus
temperature (filled, temperature sweep; open, sweeps at 8--15~K), with
Eq.~(\ref{eq:H}) for $a = 0$: fitted to 2.67 and 8--10.5~K (solid), with the
EPR $g$ and the gap that reproduces 2.67~K (grey), and for DM alone (dashed);
dotted, free-spin frequency.
(c)~$\Delta\chi^2$ of that fit versus $\Delta_0$ for three ground states (for the Y/V states scaled by $\chi^2/\nu = 1.9$), with the DM-only gap and the single-crystal range, $\Delta_0 = 98.5$--104~GHz.
All calculated positions are as the fit returns them, fit bias included (Supplementary Material, Section~S6.1).}
\end{figure*}

\section{Spin-wave analysis}\label{sec:analysis}

\subsection{Model}\label{sec:model}

With the DM vector $\bm{D}\parallel c$ we take
\begin{equation}
\begin{split}
\mathcal{H} ={}& \sum_{\langle ij\rangle_c}
\left[-J\,\bm{S}_i\!\cdot\!\bm{S}_j + \bm{D}\!\cdot\!(\bm{S}_i\times\bm{S}_j)
- d\,(S_i^xS_j^x + S_i^yS_j^y)\right]\\
&+ J'\!\!\sum_{\langle ij\rangle_{ab}}\!\bm{S}_i\!\cdot\!\bm{S}_j
- \mu_B\sum_i \bm{B}\!\cdot\!\overleftrightarrow{g}\!\cdot\!\bm{S}_i ,
\end{split}
\label{eq:H}
\end{equation}
with $S = 1/2$, ferromagnetic $J = 56$~K and $J' = 9.8$~K \cite{Tanaka1992},
the latter from the saturation field for $\bm{B}\parallel c$, 31~T
\cite{Nojiri1988}, and an easy-plane exchange anisotropy $d$ on the chain
bonds. Rotating spin $n$ of each chain by $qn$ about $c$, with $\tan q = D/J$,
removes the helix exactly and turns the DM term into an easy-plane anisotropy
$\Delta_{\rm DM} = J(\sec q - 1) = 0.22$~K at the measured pitch
$q = 5.1^\circ$ \cite{Adachi1980}, so that the total is $K = \Delta_{\rm DM} +
d$. The gapped zone-center modes are rockings of the $120^\circ$ triad out of
the easy plane, pendula whose stiffness is the anisotropy and whose inertia
comes from the exchange, through the susceptibility $(g\mu_B)^2/9J'$ of the
triangular lattice (Supplementary Material, Section~S1).
Uniform radiation, however, varies as $e^{\pm iqn}$ in the rotating frame, so
the rocking it excites advances in phase along the chain, and the chain
exchange adds $J(1 - \cos q) \simeq \Delta_{\rm DM}$ to the stiffness
\cite{Tanaka1992}. The gap is the geometric mean of the stiffness and the
exchange $9J'$ that sets the inertia,
\begin{equation}
h\Delta_0 = 3S\sqrt{J'\,[K + J(1 - \cos q)]} ,
\label{eq:gap}
\end{equation}
the exchange enhancement familiar from antiferromagnetic resonance
\cite{Keffer1952}. For DM alone the gap is 66~GHz, although $\Delta_{\rm DM}$
itself corresponds to only 4.6~GHz. A field along $c$ couples the two rockings
gyroscopically into the chiral pair of Fig.~\ref{fig:one}(b)
\cite{Tanaka1992},
\begin{equation}
\omega_\pm = \sqrt{\Delta^2 + (\gamma B/2)^2} \;\pm\; \gamma B/2,
\qquad \gamma = g_c\mu_B/h ,
\label{eq:branches}
\end{equation}
split by $\gamma B$ with $\omega_+\omega_- = \Delta^2$. At finite temperature
the exchange and anisotropy fields in Eq.~(\ref{eq:gap}) shrink with the
ordered moment, $S \to m(T)S$, so $\Delta(T) = m(T)\Delta_0$, while the Zeeman
term acts on each spin directly and is not reduced. We use
$m(T) = (1 - T/T_N)^{1/4}$, which over 8--10.5~K, where most data lie, reduces
the DM-only and single-crystal gaps to 24--46 and 36--74~GHz, and compare every
line with the
orientation-averaged spectrum of the three-sublattice model \cite{Palme1990}
at the temperature of the measurement, with the anisotropy that reproduces
Eq.~(\ref{eq:gap}), as detailed in Section~S5 of the Supplementary Material.
The $g$ tensor, $g_c = 2.08$--2.09 from EPR and
$g_{ab} - g_c = 0.042$ from magnetic torque \cite{Tanaka1985}, enters as one
scale factor; its powder average, $\langle g\rangle = 2.11$, agrees with the
Curie constants \cite{Tazuke1981}.

\subsection{Field locking}\label{sec:locking}

A crystallite precesses about its own $c$ axis only while the anisotropy pins the precession axis, and in an antiferromagnet the measure of that pinning is the
exchange-enhanced gap, not the bare anisotropy. At 10~K, where $m = 0.51$,
mean-field scaling reduces the single-crystal gap of Section~\ref{sec:gap} to
$\approx\!50$~GHz, and the closed-form reading of the field sweeps in that section
gives $81 \pm 3$~GHz for the circular runs (Supplementary Material, Table~S4), while
$\gamma B \simeq 160$--205~GHz over 5.5--7~T: the Zeeman frequency is two to
four times larger, and the precession axis should be set by $\bm{B}$. Solving
Eq.~(\ref{eq:H}) for every grain orientation, with the in-plane field taken as
uniform along the helix (Section~\ref{sec:gap}), and projecting onto the circular
basis about $\bm{k}\parallel\bm{B}$ confirms that, at 10~K and 4.5--7~T, for any
$\Delta_0$ between 66 and 130~GHz and any of the ground states of
Section~\ref{sec:gap}, the
dark channel carries less than 0.6\% of the absorption above 0.105~THz, and at
most 1.2\% for $\Delta(10~\mathrm{K}) = 81 \pm 3$~GHz (Supplementary Material, Section~S2).

\subsection{The gap}\label{sec:gap}

On single crystals the gap has been measured directly by zero-field millimeter-wave resonance, 98.5~GHz at 2~K \cite{Palme1990}, consistent with pulsed-field ESR \cite{Schmidt1998}; Eq.~(\ref{eq:gap}) then requires a two-ion anisotropy $d \approx 0.6$~K (for $S = 1/2$ a single-ion term is a constant \cite{Moriya1953}), close to the 0.48~K that Tanaka \textit{et al.} needed to fit the field dependence \cite{Tanaka1992}.

For a tilted grain our treatment takes the in-plane field as uniform along the helix, and the classical ground state is then accidentally degenerate \cite{Kawamura1985}, a one-parameter family whose members have different spectra (Supplementary Material, Section~S4); in CsCuCl$_3$ the helix and fluctuations select among them
\cite{Jacobs1993,Nikuni1998}. We treat the member with one spin along the
in-plane field ($a = 0$), the coplanar Y/V states (one spin antiparallel, or
two parallel, to it) and the family average. The cleanest handle is the 7~T line versus temperature [Fig.~\ref{fig:three}(b)]: $232.6 \pm 0.5$~GHz at 2.67~K, $220.1 \pm 0.8$~GHz at 9.9~K and, in a second run in 0.5~K steps, 225.6 to 219.2~GHz over 8--10.5~K. Fitted to the 2.67 and 8--10.5~K points with $\Delta_0$ and one $g$ [Fig.~\ref{fig:three}(c)], Eq.~(\ref{eq:H}) gives $\Delta_0 = 91.1 \pm 2.5$, $79.1 \pm 2.0$ and $70.5 \pm 2.5$~GHz for $a = 0$, the family average and the Y/V states, with
$\chi^2 = 4.2$--9.3 for five degrees of freedom and systematic errors of up to 4.4~GHz from the scaling and the referencing (Supplementary Material, Section~S4 and Table~S2), and excludes DM alone for the first two ($\Delta\chi^2 = 90$ and 39) but not for the Y/V states ($\Delta\chi^2 \approx 4$, two standard deviations).
These fits need $\langle g\rangle = 2.26$, 7\% above EPR, because near $T_N$ the calculated line tends to its free-spin value and, with the EPR $g$, stays 13--16~GHz below the measured one. With the EPR $g$ the 2.67~K line alone gives $\Delta_0 = 125$, 109 and 98~GHz for the three ground states, but the calculation then falls 4--14~GHz below the data at 8--10.5~K. These three values carry a statistical error of about 1~GHz, and referencing the 2.67~K spectrum to 20~K instead of 1.5~K raises each of them by about 3~GHz (Supplementary Material, Section~S6.1 and Table~S2). The
single-crystal gap lies within the range of the two readings; without an
independent $g$ the powder does not fix the gap. Far below $T_N$ the readings
separate: at 2.67~K and 4~T the calculated line lies at 142--148~GHz with one
$g$ and 155--165~GHz with the EPR $g$ (Supplementary Material, Section~S6.5).

The field sweeps at 8--10.5~K [Fig.~\ref{fig:three}(a)] do not follow the
calculation that fits the 7~T line: at 4--4.5~T they lie 7--15~GHz above it, and
fitted alone, with one $g$ per sweep, they reach the edge of our grid,
$\Delta_0 \geq 130$~GHz for $a = 0$ (113~GHz for the Y/V states), with $\chi^2/\nu = 2$--5 (Supplementary Material, Section~S6.5). Read with the closed form
$f_0^2 = (\langle g\rangle\mu_B B/h)^2 + \Delta(T)^2$ (Supplementary Material, Section~S7), each sweep gives $\langle g\rangle = 2.09$--2.17, near EPR, and $\Delta(T)$ falling only from 95 to 68~GHz over 8--10.5~K, against 65 to 34~GHz for the 91~GHz fit: near $T_N$, where $T_N$ depends on field and a further incommensurate phase appears \cite{Werner1997,Jacobs1998}, the gap does not collapse as the scaling assumes, which favors the EPR reading.

Above $T_N$ the line continues without a break [Fig.~\ref{fig:three}(b)] and
survives to 60~K as the paramagnetic resonance, its intensity falling roughly
as $1/T$. Corrected for the fit bias, it lies 3--7\% above the free-spin
value of the EPR tensor, $\langle g\rangle\mu_B B/h = 207$~GHz (222~GHz at
11~K, 214~GHz at 20--40~K; Supplementary Material, Sections~S6.1 and S6.4). As the field is known to a few mT, the offset is not a field error; just above $T_N$ the line gives $g_{\rm eff} = 2.26$, the $\langle g\rangle$ of the ordered-phase fits. On cooling it moves up by $21.1 \pm 1.5$~GHz between 20 and 2.67~K: the same $\Delta S_z = \pm1$
transition is the paramagnetic resonance above $T_N$ and the chiral magnon
below it.

\section{Discussion}\label{sec:discussion}

CsCuCl$_3$ already shows a chiroptical effect, magnetochiral dichroism \cite{Rikken1997} at the Cu$^{2+}$ $d$--$d$ transition near 1.5~eV \cite{Nakagawa2017}. It required enantiopure crystals, grown for the purpose \cite{Kousaka2014,Kousaka2017}, and, being governed by the static chirality, averages away in a racemic powder, as do the magnon-resonant magnetochiral effects of other helimagnets \cite{Kibayashi2014,Okamura2015}. The same material yields a handedness-resolved magnetic signal from a powder. The assignment is magnetic dipole: for an ideal proper screw the spin-current polarization vanishes \cite{Tokura2014}, so no electroactive mode of the kind found in multiferroic \cite{Skiadopoulou2017} and chiral \cite{Chaix2013} magnets is expected, and the polar phase of CsCuCl$_3$ appears only between about 14 and 16~T \cite{Miyake2015}.

Field locking is not specific to CsCuCl$_3$: it needs only a Zeeman frequency above the gap at the temperature of the measurement, $\gamma B \gg \Delta(T)$, met at accessible fields in any weakly anisotropic magnet, so the handedness of magnons relative to the field could be resolved in polycrystalline samples of the many frustrated magnets for which single crystals are unavailable. Modes whose handedness is tied to the crystal, as in altermagnets, would still average out. Such measurements need the channels referred to the linear absorption of the same runs.

\section{Conclusion}\label{sec:conclusion}

In a pressed pellet of CsCuCl$_3$, whose grains carry both structural handednesses and all orientations, the only magnetic resonance up to 1.05~THz moves up with field and absorbs one circular polarization more than 20 times as strongly as the other from 5 to 7~T, where precession about each grain's axis would give a factor of 7. The Zeeman frequency, two to four times the exchange-enhanced gap, sets the precession axis in every grain. The 7~T line gives a gap of 70--91~GHz, depending on the ground state, if one $g$, 7\% above EPR, holds from 2.67~K to $T_N$, and 98--125~GHz with the EPR $g$ at 2.67~K. Above $T_N$ the same transition continues as the paramagnetic resonance. Magnon chirality in antiferromagnets is thus accessible without a single crystal.

\section*{CRediT authorship contribution statement}

\noindent 
\textbf{J.S.R.-H.} Investigation, Resources, Writing -- review \& editing.
\textbf{G.F.F.} Investigation, Formal analysis, Writing -- review \& editing. 
\textbf{N.M.K.} Investigation, Writing -- review \& editing.
\textbf{R.S.F.} Resources, Funding acquisition, Writing -- review \& editing.
\textbf{F.G.G.H.} Formal analysis, Resources, Funding acquisition, Supervision, Project administration, Writing -- original draft, Writing -- review \& editing.

\section*{Declaration of competing interest}

The authors declare that they have no known competing financial interests or
personal relationships that could have appeared to influence the work reported
in this paper.

\section*{Declaration of generative AI and AI-assisted technologies in the
manuscript preparation process}

During the preparation of this work the authors used Claude (Anthropic) in order to edit the text and write the data-analysis code. After using this tool, the authors reviewed and edited the content as needed and take full responsibility for the content of the published article.

\section*{Acknowledgments}

\begin{sloppypar}
We acknowledge financial support from the S\~{a}o Paulo Research Foundation (FAPESP) under Grant Nos.~2021/12470-8, 2023/04245-0, 2023/16742-8, and 2026/15952-7. 
F.G.G.H. and R.S.F. acknowledge financial support from the National Council for Scientific and Technological Development (CNPq) under Grant Nos.~306550/2023-7 and 302689/2025-7, respectively. J.S.R.-H. acknowledges financial support from FAPESP Grant No.~2026/30234-3. G.F.F. acknowledges financial support from CNPq. Additional support from the INCT Advanced Quantum Materials project funded by CNPq (Grant No.~408766/2024-7), FAPESP (Grant No.~2025/27091-3), and CAPES is also gratefully acknowledged.
\end{sloppypar}

\section*{Data availability}

Data will be made available on request.

\section*{Supplementary material}

The Supplementary Material derives the model used here and documents the sample and the processing and analysis of the data, with their checks and uncertainties.

\onecolumn
\setcounter{section}{0}
\setcounter{subsection}{0}
\setcounter{equation}{0}
\setcounter{figure}{0}
\setcounter{table}{0}
\renewcommand{\thesection}{S\arabic{section}}
\renewcommand{\thesubsection}{\thesection.\arabic{subsection}}
\renewcommand{\theequation}{S\arabic{equation}}
\renewcommand{\thefigure}{S\arabic{figure}}
\renewcommand{\thetable}{S\arabic{table}}
\renewcommand{\theHsection}{S\arabic{section}}
\renewcommand{\theHsubsection}{S\arabic{section}.\arabic{subsection}}
\renewcommand{\theHequation}{S\arabic{equation}}
\renewcommand{\theHfigure}{S\arabic{figure}}
\renewcommand{\theHtable}{S\arabic{table}}
\renewcommand{\citenumfont}[1]{S#1}
\renewcommand{\bibnumfmt}[1]{[S#1]}
\renewcommand{\refname}{Supplementary references}

\begin{center}
{\Large\bfseries Supplementary Material}\\[6pt]
{\large Terahertz detection of magnon chirality in polycrystalline CsCuCl$_3$}
\end{center}
\vspace{2ex}

Equation and figure numbers without an ``S'' refer to the main text.
Frequencies are quoted in GHz with $1~\mathrm{K} = 20.84$~GHz, and
$\gamma = g\mu_B/h = 13.996\,g$~GHz/T.

\section{Rotating frame and the zero-field gap}\label{sm:sec:gap}

\textit{Rotating frame.} For two neighboring spins in the $ab$ plane rotated by
$q$ about $c$, the chain bond of Eq.~(2) of the main text has the energy
$E(q) = -S^2(J\cos q + D\sin q)$, minimal at $\tan q = D/J$; the measured pitch
$q = 5.1^\circ$ per Cu \cite{Adachi1980S} fixes $D = J\tan q = 5.0$~K. Writing
$\bm{S}_n = R_z(qn)\tilde{\bm{S}}_n$ removes the twist exactly. The antisymmetric
part of the bond vanishes and the bond matrix becomes
$\mathrm{diag}(-\sqrt{J^2+D^2} - d,\,-\sqrt{J^2+D^2} - d,\,-J)$. A pair of parallel spins along the chain gains
\begin{equation}
K = \Delta_{\rm DM} + d, \qquad
\Delta_{\rm DM} = \sqrt{J^2+D^2} - J = J(\sec q - 1) = 0.2226~\mathrm{K},
\label{sm:eq:K}
\end{equation}
by lying in the $ab$ plane rather than along $c$. For rigid chains, i.e.\ for rotating-frame modes at chain wavevector $k = 0$, the intrachain bonds act only through this easy-plane anisotropy $K(S^z)^2$ per spin. One $D$ sets the pitch, $\Delta_{\rm DM}$ and the term $J(1 - \cos q)$ below, so the DM-only model ($d = 0$) has no free parameter. In this frame an in-plane field component rotates by $q$ from site to site along the chain; we treat it as uniform (Sec.~\ref{sm:sec:degeneracy}).

\textit{Rigid rocking.} At $B = 0$ the triad lies in the easy plane. The two gapped modes are rockings of the triad about an in-plane axis, which tip the spin plane out of the easy plane; the rotation about $c$ is the Goldstone mode. For a rocking angle $\varphi$ the three spins acquire, on average, $(S^z)^2 = S^2\varphi^2/2$, so a rigid rocking costs $\tfrac12 KS^2\varphi^2$ per spin. The inertia comes from the exchange: a rotating antiferromagnet cants and develops a moment as if it were in a field $\hbar\dot\varphi/(g\mu_B)$ (Larmor's theorem), which costs $\tfrac12\chi(\hbar\dot\varphi/g\mu_B)^2$ per spin. For the triangular antiferromagnet $\chi = (g\mu_B)^2/(9J')$, because the magnetization rises linearly to saturation \cite{Andreev1980S}. The rigid rocking is therefore a pendulum of stiffness $KS^2$ and moment of inertia $\hbar^2/(9J')$, with angular frequency $3S\sqrt{J'K}/\hbar$.

\textit{Observable modes.} Radiation is uniform in the laboratory frame, not in the rotating one. Its component along $c$ is unchanged by $R_z$ and couples to rigid modes, but its circular in-plane components acquire the phase $e^{\mp iqn}$ on spin $n$ and therefore excite the rotating-frame modes at $k = \pm q$ \cite{Tanaka1992S}. In these the rocking advances in phase by $q$ from spin to spin along the chain, and the out-of-plane components of neighboring spins, coupled by $-J$ in the rotating frame, add $\tfrac12 J(1 - \cos q)S^2\varphi^2$ per spin to the restoring energy; the
corresponding cost of the in-plane components, of the same size, is negligible
against the $9J'$ that sets the inertia. The stiffness of the observable rocking
is thus $K + J(1 - \cos q)$, and
\begin{equation}
h\Delta_0 = 3S\sqrt{J'\,[K + J(1 - \cos q)]} ,
\label{sm:eq:gapS}
\end{equation}
Eq.~(3) of the main text, with $\Delta_0$ a frequency. This is the zero-field limit of the mean-field resonance condition of Tanaka \textit{et al.} [Eq.~(18) of Ref.~\cite{Tanaka1992S}], $\omega_0/\gamma = [3H_{E1}(H_{\rm DM}^2/H_{E0} + H_A)]^{1/2}$, written in exchange constants: $g\mu_B H_{E1} = 3J'S$, $g\mu_B H_{\rm DM}^2/H_{E0} = (D^2/J)S$, which equals $[\Delta_{\rm DM} + J(1 - \cos q)]S$ to order $q^2$, and $g\mu_B H_A = dS$. Since $J(1 - \cos q) = 0.2217$~K $\simeq \Delta_{\rm DM}$, the helix doubles the stiffness of DM origin. DM alone, $\Delta_{\rm DM} = 0.2226$~K or 4.6~GHz, gives $\Delta_0 = 1.5\sqrt{9.8 \times 0.4443}~\mathrm{K} = 3.130~\mathrm{K} = 65.2$~GHz, where rigid chains would give 46.2~GHz. Linear spin-wave theory at $k = q$ (Sec.~\ref{sm:sec:lswt}) gives 65.5~GHz and agrees with Eq.~(\ref{sm:eq:gapS}) to 0.5\% for every $d$. The analysis uses the grid $\Delta_0 = 65.5$ (DM only), 72.5, 80, 88, 95, 101, 110, 120 and 130~GHz, set by $d$.

\textit{The single-crystal gap.} On single crystals the zone-center gap has been measured by zero-field millimeter-wave resonance, 98.5~GHz at 2~K~\cite{Palme1990S}. With the temperature scaling of Sec.~\ref{sm:sec:degeneracy}, $m(2~\mathrm{K}) = 0.95$, this corresponds to $\Delta_0 = 104$~GHz, or to 98.5~GHz if the moment is already saturated at 2~K; we quote $\Delta_0 = 98.5$--104~GHz and use the grid point 101~GHz as the reference. Spin-wave theory reaches it at $d = 0.56$--0.67~K, i.e.\ $K = 0.78$--0.90~K. Tanaka \textit{et al.} fitted the frequency--field diagram of Palme \textit{et al.} for $\bm{B}\parallel c$ with $H_A = 0.18$~T, i.e.\ $d = 0.48$~K, which gives 94.7~GHz in spin-wave theory with our parameters~\cite{Tanaka1992S}. The antiferromagnetic-resonance model fitted to pulsed-field ESR, with exchange field $B_{\rm ex} = 10$~T and anisotropy field $B_a = 0.36$~T~\cite{Schmidt1998S}, gives $\sqrt{3B_aB_{\rm ex}} = 3.29$~T, i.e.\ 97.5--98.4~GHz for $g = 2.12$--2.14, consistent with it.

\textit{The $g$ tensor.} The calculations use $J = 56$~K and $J' = 9.8$~K~\cite{Tanaka1992S} and the tensor of the Curie constants below the Jahn--Teller transition, $C_\parallel = 0.43 \pm 0.01$ and $C_\perp = 0.42 \pm 0.01$~emu\,K/mol~\cite{Tazuke1981S}, i.e.\ $(g_c, g_{ab}) = (2.14, 2.12)$, each $\pm 0.03$, and $\langle g\rangle = 2.12 \pm 0.02$, consistent with EPR and torque, $g_c =
2.08$--2.09 and $\langle g\rangle = 2.11$~\cite{Tanaka1985S}. For the powder, replacing one tensor by the other is equivalent to a 0.6\% change of the field scale, with residuals within 0.03~GHz over 5.25--7~T at 8~K. In fits the tensor is multiplied by one scale factor, and we quote the fitted powder average $\langle g\rangle$, which does not depend on this choice. The free-spin frequency of the EPR tensor at 7~T is $\langle g\rangle\mu_B B/h = 207.0$~GHz.

\section{Field dependence and the locking of the precession axis}\label{sm:sec:field}

\textit{Field along $c$.} The field couples the two rockings through a gyroscopic term. With $\varphi = \varphi_x + i\varphi_y$,
\begin{equation}
\ddot\varphi + \Delta^2\varphi + i\gamma B\,\dot\varphi = 0
\;\Rightarrow\;
\omega_\pm = \sqrt{\Delta^2 + (\gamma B/2)^2} \pm \gamma B/2 ,
\qquad \omega_+\omega_- = \Delta^2,\quad \omega_+ - \omega_- = \gamma B ,
\label{sm:eq:branchesS}
\end{equation}
which is Eq.~(4) of the main text: the two linear rockings become circular motions of opposite sense, split by $\gamma B$. At high field $\omega_+ \simeq \gamma B + \Delta^2/\gamma B$ approaches the free-spin line, and $\omega_- \simeq \Delta^2/\gamma B$ softens. With $\bm{D}\parallel c$ and $J'$ isotropic, $S_z$ is conserved for $\bm{B}\parallel c$ and the modes carry $\Delta S_z = \pm1$.

\textit{Field in the easy plane.} The rocking about the field direction has its oscillating moment parallel to $\bm{B}$, so light travelling along $\bm{B}$ cannot excite it. The other rocking couples to the rotation about $c$ and moves to $\sqrt{\Delta^2 + (\gamma B)^2} \simeq \gamma B + \Delta^2/2\gamma B$; spin-wave theory brackets this closed form, by an amount that depends on the classical ground state (Sec.~\ref{sm:sec:degeneracy}).

\textit{Locking.} In both geometries the active mode ends up precessing about $\bm{B}$ in the Larmor sense, and the calculation of Sec.~\ref{sm:sec:lswt} shows how complete this is for arbitrary grain orientations. At 10~K, for $\Delta_0 = 101$~GHz [$\Delta(10~\mathrm{K}) = 51$~GHz] and $a = 0$, the fraction of the absorption of the upper gapped mode, the one observed, that falls in the dark helicity reaches 0.15 at 0.5~T ($\gamma B/\Delta = 0.3$) for grains with $c\perp\bm{B}$, 0.012 at 2~T, and stays below $10^{-4}$ at 7~T for every angle between $c$ and $\bm{B}$; averaged over a random powder the dark-to-linear ratio is 0.04, 0.008 and $3\times10^{-5}$ at 0.5, 2 and 7~T. At 10~K, the temperature of the circular measurement, over 4.5--7~T, for gaps of 66--130~GHz and every choice of ground state (Sec.~\ref{sm:sec:degeneracy}), the dark channel carries less than 0.6\% of the absorption above 0.105~THz (0.2\% for 101~GHz); at 2.67~K, where the gap is largest, it reaches 1\% at 101~GHz and 2\% at 130~GHz. The closed form of Sec.~\ref{sm:sec:closed} reads the field sweep of the circular runs as
$\Delta(10~\mathrm{K}) = 81 \pm 3$~GHz (Table~\ref{sm:tab:closedsweeps}), above the
33--66~GHz that the scaling gives at 10~K for the grid. For that gap at 10~K, i.e.\
$\Delta_0 = 154$--166~GHz with the scaling, the dark channel carries at most 1.2\%, at
4.5~T for the Y/V states (0.1\% for $a = 0$). The locking therefore holds with either
reading of the gap, for which $\gamma B/\Delta(10~\mathrm{K}) = 2$--4 over 5.5--7~T with $\gamma = g_c\mu_B/h$. Precession about each grain's own $c$ axis would instead leave a dark-to-linear ratio of 1/4 (Sec.~\ref{sm:sec:powder}).

\section{Powder selection rules: 7/4 and 1/4 versus 2 and 0}\label{sm:sec:powder}

Light travels along $\hat{\bm{z}}\parallel\bm{B}$. A mode is absorbed through its oscillating moment, a complex vector $\bm{m}$; the helicities $\bm{e}_\pm = (\hat{\bm{x}} \pm i\hat{\bm{y}})/\sqrt{2}$ absorb $A_\pm = |\bm{e}_\pm^*\!\cdot\bm{m}|^2$, and linearly polarized light, an equal mix of the two, absorbs $A_{\rm lin} = (A_+ + A_-)/2$. A moment circling in the $xy$ plane, $\bm{m} = \hat{\bm{x}} + i\hat{\bm{y}}$, gives $A_+ = 2$, $A_- = 0$ and $A_{\rm lin} = 1$: the channel ratios of a mode that precesses about $\bm{B}$ are 2 and 0, in every grain.

If instead each grain precessed about its own $c$ axis, tilted by $\theta$ from $\bm{k}$, its circle would appear to the light foreshortened into an ellipse,
$\bm{m} = \cos\theta\,\hat{\bm{x}} + i\hat{\bm{y}}$, with
\begin{equation}
A_\pm(\theta) = \frac{(1 \pm |\cos\theta|)^2}{2} ;
\label{sm:eq:Apm}
\end{equation}
the absolute value appears because the sense of precession about $c$ is fixed by the sign of $\bm{B}\cdot c$. For random orientations $u = |\cos\theta|$ is uniform on $[0,1]$, since the solid angle $2\pi\sin\theta\,d\theta = 2\pi\,|d\cos\theta|$, so
\begin{equation}
\langle A_+\rangle = \int_0^1\frac{(1+u)^2}{2}\,du = \frac{7}{6},\qquad
\langle A_-\rangle = \int_0^1\frac{(1-u)^2}{2}\,du = \frac{1}{6},\qquad
A_{\rm lin} = \frac{2}{3},
\label{sm:eq:powder}
\end{equation}
and the channel ratios are $7/4 = 1.75$ and $1/4 = 0.25$, with a contrast $(A_+ - A_-)/(A_+ + A_-) = 3/4$. Passed through our processing, with the measured line centers, the $x$-polarized absorption of the same runs and the optical thickness of the pellet, precession about $\bm{B}$ gives a half-sum and a half-difference of the channels over the $x$-polarized absorption of 1.02--1.08 each over 4.5--7~T, and grain locking 1.01--1.04 and 0.76--0.79 (Sec.~\ref{sm:sec:estimators}). Pressed needles tend to lie with $c$ in the pellet plane, perpendicular to $\bm{k}$, which pulls the grain-locked ratios toward 1 and 1.

\section{Ground-state degeneracy, temperature scaling and the gap}\label{sm:sec:degeneracy}

\textit{Degeneracy.} With rigid chains in the rotating frame, the interchain energy
of the three sublattices depends only on their total spin $\bm{S}_{\rm tot}$, as $(3J'/2)|\bm{S}_{\rm tot}|^2$ per triad up to a constant; the Zeeman energy depends only on $\bm{S}_{\rm tot}$; and the chain bonds contribute $K\sum_i (S_i^z)^2$. For given $\bm{S}_{\rm tot}$ the last term is minimized by equal $c$ components, which
leaves three in-plane projections of equal length $\ell$ that must add up to the
in-plane part of $\bm{S}_{\rm tot}$, of length $\Lambda$. For any field with a
component in the $ab$ plane this is a one-parameter family of states of identical
energy, an accidental degeneracy of the kind found in the Heisenberg triangular
antiferromagnet in a field \cite{Kawamura1985S}, reduced here to one parameter by
the easy-plane anisotropy; only for $\bm{B}\parallel c$ is the ground state
unique. We parametrize it by the angle $a$ of one projection with respect to the
in-plane total: $a = 0$ puts one spin along the in-plane field; the opposite
extreme is the coplanar Y state, one spin antiparallel ($a = 180^\circ$), which
exists while $\Lambda \le \ell$, i.e.\ below one third of the in-plane saturation
moment, and the V state (two spins parallel) above it, at
$\cos a = (\Lambda^2 - 3\ell^2)/(2\Lambda\ell)$. The members have different
spectra. For $\bm{B}\perp c$ at 7~T and $\Delta_0 = 101$~GHz the active mode lies at
224.3~GHz for $a = 0$ and 244.0~GHz for the Y state, and grains at intermediate
orientations differ by less. Which member is realized is decided by physics outside
Eq.~(2) of the main text: fluctuations favor the coplanar states in the triangular
antiferromagnet \cite{Kawamura1985S,Chubukov1991S}, and in CsCuCl$_3$ the selection in
transverse fields has been analyzed together with the helix
\cite{Jacobs1993S,Nikuni1998S}; the helix, along which the in-plane field direction
rotates relative to the triad, sweeps the member along the chain, which for the
uniform response amounts to an average over the family. We therefore repeat the
analysis for three choices: $a = 0$, the equal-weight average over the family ($a$
from 0 to its maximum in quarters), and the Y/V extreme. The dark-channel fraction
of Sec.~\ref{sm:sec:field} does not depend on the choice.

\textit{Temperature scaling.} Both ingredients of Eq.~(\ref{sm:eq:gapS}) are molecular
fields, which in mean-field theory shrink with the ordered moment, $S \to m(T)S$, so
that $\Delta(T) = m(T)\Delta_0$. The calculation multiplies $J$, $D$, $J'$ and the
anisotropy by $m(T)$, which leaves the pitch unchanged; the Zeeman term is not scaled,
because the applied field acts on each spin directly, so at fixed high field the line
approaches the free-spin value as $T\to T_N$. We use $m(T) = (1 - T/T_N)^\beta$ with
$T_N = 10.7$~K \cite{Adachi1980S} (Ref.~\cite{Palme1990S} gives 10.5~K) and
$\beta = 1/4$, i.e.\ $m = 0.93$, 0.71, 0.63, 0.51 and 0.37 at 2.67, 8, 9, 10 and
10.5~K. The power law is a near-$T_N$ form used over the whole range; $T_N \pm 0.2$~K
and $\beta = 0.20$--0.35 move the gap estimate of this section by up to 3.6~GHz for each ground state (Table~\ref{sm:tab:scaling}). Two effects are not included: $T_N$ depends on field \cite{Werner1997S}, and for in-plane fields above 5~T a further incommensurate phase occupies a pocket of the phase diagram just below $T_N$
\cite{Werner1997S,Jacobs1998S}; both matter most at 0.75--0.98~$T_N$, where the low-temperature points and the field sweeps lie.

\textit{Gap estimate.} Every calculated line is passed through the processing of the
data (Sec.~\ref{sm:sec:processing}) and compared with the fitted centers, with one scale
factor of the $g$ tensor per fit on a grid equivalent to $\langle g\rangle = 1.95$--2.34. $\Delta_0$, its error and $\chi^2_{\rm min}$ come from a parabola through the lowest point of the grid of $\Delta_0$ (Sec.~\ref{sm:sec:gap}) and its two neighbours; $\Delta\chi^2$ is measured from $\chi^2_{\rm min}$, errors and $\Delta\chi^2$ are scaled by $\chi^2/\nu$ where it exceeds 1, and $\langle g\rangle$ is
taken at the fitted $\Delta_0$. The estimator of the main text, S$^+$, fits the 7~T
line at 2.67~K, the lowest temperature of the 7~T temperature sweep, and at the six
points of the low-temperature sweep below $T_N$, 8--10.5~K (Table~\ref{sm:tab:closed}),
with $\Delta_0$ and one $g$. It gives $91.1 \pm 2.5$, $79.1 \pm 2.0$ and $70.5 \pm 2.5$~GHz for $a = 0$, the family average and the Y/V states, with $\chi^2/\nu = 0.8$--1.9 (Table~\ref{sm:tab:degeneracy}); the temperature scaling and the zero-field reference of the 2.67~K spectrum move it by up to 4.4~GHz (Table~\ref{sm:tab:scaling}). DM alone is excluded for $a = 0$ and the family average ($\Delta\chi^2 = 90$ and 39) but not for the Y/V states. The fits need $\langle g\rangle = 2.26$, 7\% above EPR: at 10.5~K the calculated line, processed like the data, lies within 4~GHz of its
free-spin value for any gap, and with the EPR $g$ 13--16~GHz below the measured one,
which, corrected for the fit bias, lies 16~GHz above the free-spin value of the EPR
tensor (Table~\ref{sm:tab:closed}; Sec.~\ref{sm:sec:para}); only a larger $g$ reaches it, and the 2.67~K point then fixes the gap. With the EPR $g$ the 2.67~K line alone requires 125, 109 and 98~GHz (Table~\ref{sm:tab:scaling}), and the calculation with mean-field scaling then falls 4--14~GHz below the data at 8--10.5~K. Fitted alone, the field sweeps give larger gaps (Table~\ref{sm:tab:degeneracy}) but disagree with the calculation at the temperatures of the low-temperature points (Sec.~\ref{sm:sec:sweeps}), and are not used for the gap.

\begin{table}[h]
\caption{\label{sm:tab:degeneracy}Zero-temperature gap (GHz) for three choices within the
degenerate ground state, with $\Delta\chi^2$ of the DM-only gap (65.5~GHz) and of the
single-crystal gap (101~GHz) in parentheses. Forward model (Sec.~\ref{sm:sec:processing})
at $\sigma_t = 10$~ps, Lorentzian line shape, intrinsic FWHM 10~GHz, 13 orientations.
Last column: $\chi^2/\nu$ for $a = 0$, family average and Y/V. ``$\geq 130$'': minimum
at the upper edge of the grid, with the vertex of the parabola outside it.}
\centering
\begin{tabular}{lcccc}
\hline\hline
Estimator & $a = 0$ & family average & Y/V states & $\chi^2/\nu$\\
\hline
7~T, 2.67 and 8--10.5~K (S$^+$) & $91.1 \pm 2.5$ (90/19) & $79.1 \pm 2.0$ (39/139) & $70.5 \pm 2.5$ (4.1/186) & 0.8, 0.9, 1.9\\
Field sweeps, $g$ per run & $\geq 130$ & $122.8 \pm 2.2$ & $113.3 \pm 1.9$ & 4.9, 2.7, 2.2\\
\hline
$\langle g\rangle$ of the S$^+$ fit & 2.259 & 2.259 & 2.256 & \\
\hline\hline
\end{tabular}
\end{table}

\begin{table}[h]
\caption{\label{sm:tab:scaling}The S$^+$ estimate (GHz) under changes of the temperature
scaling $m(T) = (1 - T/T_N)^\beta$ and of the zero-field reference of the 2.67~K
spectrum, with $\Delta\chi^2$ of the DM-only gap in parentheses; and the gap for which
the calculation reproduces the 2.67~K line alone at a fixed $g$. The model depends on
temperature only through $m$ and is interpolated in $m$ on a fine temperature grid
(1.5--10.5~K), which reproduces the S$^+$ row of Table~\ref{sm:tab:degeneracy} within 0.1~GHz in $\Delta_0$; the interpolation changes its $\Delta\chi^2$ for DM alone from 90, 39 and 4.1 to 90, 39 and 3.8. For $T_N = 10.5$~K the 10.5~K point is left out. The fixed-$g$ gaps are those at which the calculated line crosses the measured one; they have statistical errors of 0.8--1.2~GHz (1.3--1.8~GHz referenced to 20~K), and referencing the 2.67~K spectrum to 20~K, above $T_N$, instead of 1.5~K raises them by 2.6--3.4~GHz.}

\centering
\begin{tabular}{lccc}
\hline\hline
 & $a = 0$ & family average & Y/V states\\
\hline
$T_N = 10.7$~K, $\beta = 1/4$ & $91.1 \pm 2.5$ (90) & $79.2 \pm 2.0$ (39) & $70.5 \pm 2.6$ (3.8)\\
$T_N = 10.9$~K & $93.3 \pm 2.4$ (100) & $81.4 \pm 2.2$ (48) & $72.8 \pm 2.8$ (6.8)\\
$T_N = 10.5$~K & $90.8 \pm 2.7$ (75) & $79.2 \pm 2.2$ (34) & $71.1 \pm 2.1$ (7.2)\\
$\beta = 0.20$ & $94.5 \pm 3.1$ (67) & $82.7 \pm 3.7$ (20) & $74.2 \pm 4.8$ (3.2)\\
$\beta = 0.30$ & $88.9 \pm 2.4$ (81) & $77.4 \pm 1.9$ (32) & $69.0 \pm 1.7$ (4.1)\\
$\beta = 0.35$ & $87.7 \pm 2.3$ (73) & $76.3 \pm 1.8$ (27) & $68.0 \pm 1.6$ (2.4)\\
2.67~K referenced to 20~K & $95.5 \pm 3.5$ (56) & $82.0 \pm 3.6$ (19) & $71.4 \pm 4.2$ (1.9)\\
\hline
2.67~K alone, EPR $g$ ($\langle g\rangle = 2.113$) & 125 & 109 & 98\\
\quad the same, referenced to 20~K & 128 & 112 & 101\\
2.67~K alone, $\langle g\rangle = 2.186$ & 110 & 96 & 86\\
\hline\hline
\end{tabular}
\end{table}

\begin{table}[h]
\caption{\label{sm:tab:closed}The 7~T line used in the gap estimate of the main text (S$^+$):
fitted center and center corrected for the fit bias by inverting the forward model of a
single line (Sec.~\ref{sm:sec:processing}). The 2.67~K spectrum is referenced to the 1.5~K scan.}
\centering
\begin{tabular}{lccccccc}
\hline\hline
$T$ (K) & 2.67 & 8.00 & 8.49 & 9.01 & 9.49 & 10.01 & 10.50\\
fitted (GHz) & $232.6 \pm 0.5$ & $225.6 \pm 0.6$ & $224.0 \pm 0.7$ & $222.3 \pm 0.8$ & $220.9 \pm 0.8$ & $220.1 \pm 0.8$ & $219.2 \pm 0.9$\\
corrected (GHz) & 236.2 & 229.5 & 227.9 & 226.2 & 224.8 & 224.0 & 223.1\\
\hline\hline
\end{tabular}
\end{table}

\section{Linear spin-wave theory}\label{sm:sec:lswt}

The powder calculation uses the three-sublattice model with rigid chains \cite{Palme1990S}, i.e.\ the rotating-frame modes at $k = 0$, with the easy-plane anisotropy raised from $K$ to $K_{\rm eff}$ so that its zero-field gap equals that of the observable modes at $k = \pm q$; $K_{\rm eff} - K = 0.226$--0.240~K, close to $J(1 - \cos q)$. For every grain orientation and field: (i)~for a field of given strength at angle $\theta$ to $c$, the classical energy of the three sublattices is minimized from the umbrella and several random starts, and the members of the degenerate family of Sec.~\ref{sm:sec:degeneracy} are constructed from the result; (ii)~each sublattice gets a local frame with its $z$ axis along the classical spin, and the Holstein--Primakoff substitution \cite{Holstein1940S}, $S^z = S - a^\dagger a$, $S^x + iS^y \simeq \sqrt{2S}\,a$, is truncated at quadratic order, which gives the same frequencies as the linearized classical equations of motion; (iii)~the quadratic Hamiltonian is a $6\times6$ bosonic matrix, diagonalized by a Bogoliubov transformation with Colpa's method \cite{Colpa1978S}. It returns the three frequencies (the soft mode at zero and the two gapped modes) and each mode's transition moment $\bm{d}_\nu = \langle\nu|\bm{M}|0\rangle$, with $\bm{M} = \mu_B\overleftrightarrow{g}\!\cdot\!\sum_i\bm{S}_i$, from which $a_\pm = |\bm{e}_\pm^*\!\cdot\bm{d}_\nu|^2$ for light along $\bm{B}$; (iv)~the pellet average weights each orientation by $\sin\theta$, with 13 orientations for the fits, whose sampling noise on the line position, about 1~GHz, is below the errors of the line centers; every mode, with its absorption weight, is passed through the processing of the data and fitted like the data (Sec.~\ref{sm:sec:processing}), so that calculated and measured fitted centers are compared; the centroid of the upper branch, the one in the fit window, is used only in the closed form of Sec.~\ref{sm:sec:closed} (at the gaps, fields and temperatures where it is used, the lower branches lie at 17--74~GHz, below the window, and carry at most 1.8\% of the weight); (v)~at temperature $T$ the molecular fields are scaled as in Sec.~\ref{sm:sec:degeneracy}.

\textit{Checks and limits.} At $B = 0$ the gap equals that of the $k = q$ modes by construction, 0.5\% above Eq.~(\ref{sm:eq:gapS}). For $\bm{B}\parallel c$ the upper branch agrees with the full $k = q$ calculation to 0.8\% or better over 0--7~T, and with Eq.~(\ref{sm:eq:branchesS}) to 3.3\% over the grid of gaps. For $m \to 0$ the line tends to the free-spin value of the tensor used, $\langle g\rangle = 2.127$ before scaling. For tilted grains a complete treatment would have to include the coupling of the $k = 0$ and $k = \pm q$ sectors by the in-plane field, which rotates along the chain in the rotating frame. Treating the two sectors separately, each in a uniform
in-plane field, is not consistent: it leaves the upper mode strongly elliptical even
at 7~T, where the Zeeman frequency exceeds the gap two- to fourfold and the precession about $\bm{B}$ must be circular up to corrections of order $(\Delta/\gamma B)^2$, and we do not use it. The calculation omits quantum fluctuations beyond linear order, which for $S = 1/2$ are not small in general although the strong ferromagnetic chains suppress them, and magnon interactions and damping, so that it predicts no line width.

\section{Data analysis}\label{sm:sec:data}

\subsection{Sample, raw traces and processing}\label{sm:sec:processing}
\begin{figure}[t]
\centering
\includegraphics[width=0.6\textwidth]{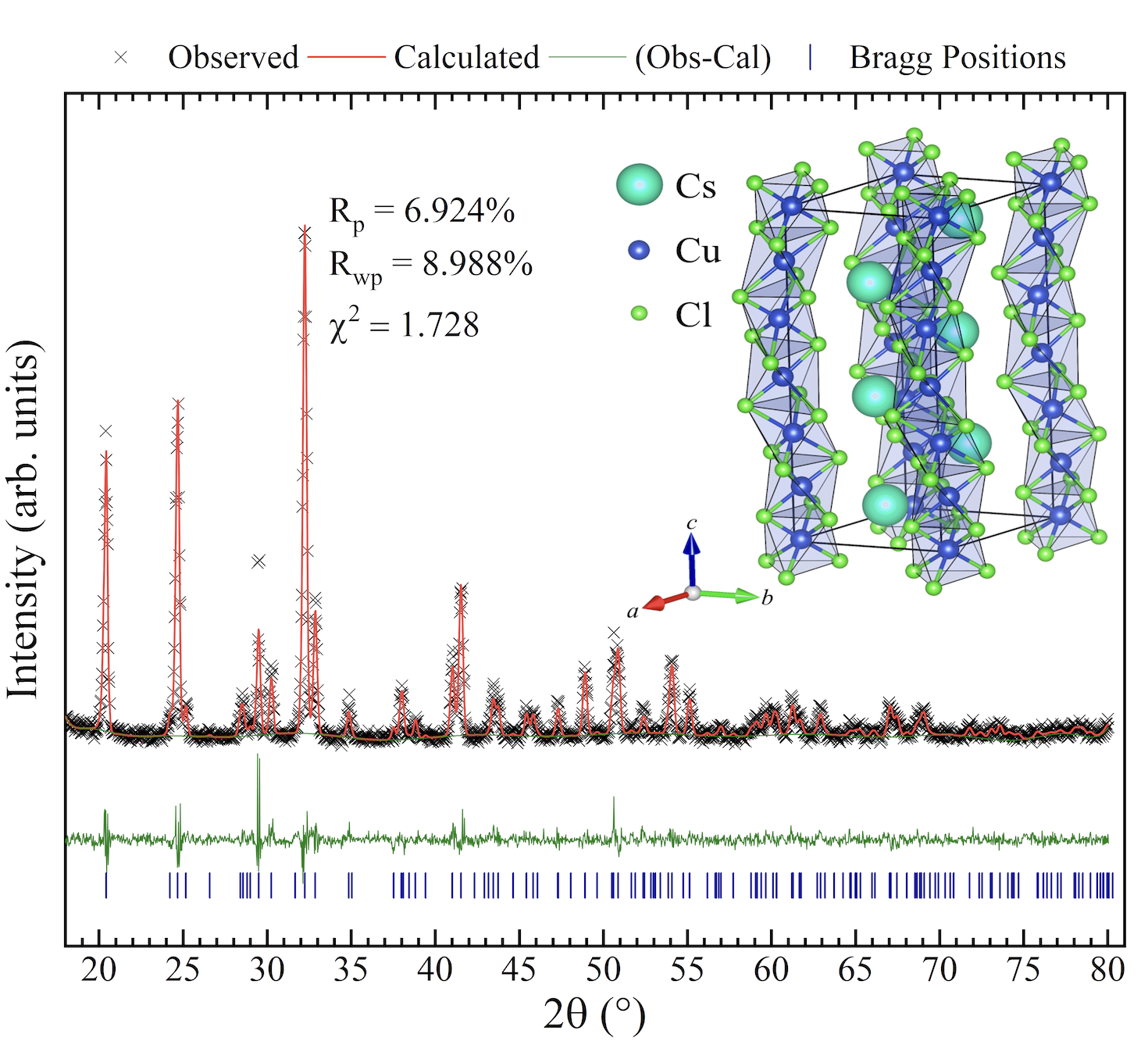}
\caption{X-ray powder diffraction of the ground crystals before pressing (crosses), with
the Rietveld fit (red line), the difference between the two (green line) and the Bragg
positions of CsCuCl$_3$, $P6_122$/$P6_522$ (ticks). Inset: crystal structure.}
\label{sm:fig:xrd}
\end{figure}

Dark red needle-shaped single crystals of CsCuCl$_3$ were grown by slow evaporation
from a solution containing stoichiometric amounts of CsCl and CuCl$_2$ in hydrochloric
acid (47~wt.~\% in H$_2$O)~\cite{RodriguezHernandez2024S,Kousaka2014S}. The solution was
heated at 120~$^\circ$C for 1~h and then kept at room temperature in a beaker sealed
with paraffin film. The crystals were collected and washed with toluene. The structure
was confirmed by single-crystal X-ray diffraction~\cite{RodriguezHernandez2024S}. The
crystals were then ground in an agate mortar, giving particles of 10--50~$\mu$m, and
pressed into a pellet of thickness $L = 1.3$~mm. The ground powder was characterized by
X-ray powder diffraction before pressing (Fig.~\ref{sm:fig:xrd}). Rietveld refinement
against the structure of Ref.~\cite{RodriguezHernandez2024S} reproduces the pattern with
$R_{\rm p} = 6.92\%$, $R_{\rm wp} = 8.99\%$ and $\chi^2 = 1.73$; every observed
reflection is indexed by CsCuCl$_3$~\cite{Schlueter1966S,Hirotsu1975S,Hirotsu1977S} and
no secondary phase is detected, confirming that the powder is single phase and
homogeneous. Powder diffraction does not distinguish between the two enantiomorphic
models, $P6_122$ and $P6_522$, which give identical patterns, so the refinement
constrains the lattice and the phase but not the chirality; the presence of both
handednesses follows instead from the growth
conditions~\cite{RodriguezHernandez2024S,Kousaka2014S,Inomata2026S}.

The pulse through the pellet arrives 6.19~ps after the pulse through the empty holder, which gives a
group index $n_g = 2.43$ and puts the internal echo of the pellet at
$2n_gL/c = 21$~ps behind the pulse; a second echo, 29~ps behind the pulse, is
present with and without the pellet and belongs to the setup.

Every spectrum is computed from the raw traces of the spectrometer (1001 points,
0.05~ps apart, 50~ps), of the pellet and of the empty holder, a 3~mm hole measured
as a temperature sweep at 0~T and interpolated linearly in the sensor temperature to
that of each pellet trace. To remove any change of the empty holder with field, $\Delta\alpha$ uses the zero-field trace of the pellet itself. Every trace is
multiplied by a Gaussian window $\exp[-(t - t_c)^2/2\sigma_t^2]$ centered on its own
peak $t_c$, zero-padded to $2^{14}$ points and Fourier transformed, and
$T(f) = |S_{\rm pellet}|^2/|S_{\rm holder}|^2$; we use $\sigma_t = 10$~ps. The window
multiplies the field, so the spectra are convolved with a Gaussian of
$\sigma_f = 1/2\pi\sigma_t = 15.9$~GHz, 37.5~GHz FWHM. It reduces the internal echo to
$\exp(-21^2/200) = 11\%$ of its amplitude and the setup echo to 1.5\%.

\textit{Referencing.}
Each spectrum is referenced to zero field at the same temperature [Eq.~(1) of the main text], which cancels the phonons (the $A_2$(TO) mode at 0.52~THz and a second one near 1.2~THz, both field independent and present at 300~K), the Fabry--P\'{e}rot structure and the instrument response; no fitted oscillator is subtracted. On the scale of the transmittance the magnon is small: at 8~K and 7~T it lowers the transmitted intensity by 22\% at the line center. The 7~T temperature sweep is referenced to the zero-field scan closest in temperature (the 0~T sweep has scans below 30~K only at 1.5, 9.86 and 20~K, so the 2.67~K spectrum is referenced to the 1.5~K scan), and the low-temperature sweeps (8--15~K in 0.5~K steps at 0 and 7~T) to each other at the same setpoint. Referencing the 7~T spectra to the 20~K scan instead moves the 2.67~K center from $232.6 \pm 0.5$ to $234.1 \pm 0.8$~GHz and the 9.9~K center from $220.1 \pm 0.8$ to $220.4 \pm 0.9$~GHz, and raises the gap estimate of the main text by 0.8--4.4~GHz (Table~\ref{sm:tab:scaling}). The zero-field absorption at 1.5~K exceeds that at 20~K by 0.77~cm$^{-1}$ at 110~GHz, falling to 0.09~cm$^{-1}$ at 200~GHz: the upper wing of the zero-field mode, at 98.5~GHz in single crystals~\cite{Palme1990S}, which the 1.5~K reference subtracts from the low-frequency side of the 2.67~K line.

\textit{Resolution function and fit bias.} To first order in $\Delta\alpha L$ the
measured absorption is the true one convolved with the Gaussian weighted by the
amplitude spectrum of the pulse, which rises steeply across the line:
$d\ln|E|/d\ln f = 2.0$--4.9 through the pellet and 2.6--4.0 through the empty holder
at 0.13--0.23~THz. Frequencies above each point therefore weigh more than those below
it, and the fitted center of a narrow line comes out low. We call this shift the fit
bias: injecting a causal Lorentzian into the measured zero-field trace and processing
and fitting it like the data gives $-3.4$ to $-4.4$~GHz at 0.17--0.23~THz, close to
$-3\sigma_f^2/f$ (intrinsic FWHM 10--25~GHz; zero-field traces of the three linear field sweeps, the circular runs and the low-temperature sweep). The fit bias cancels wherever a calculated line is compared with the data, because the calculation goes through the same processing (forward model, below), so the gap fits need no correction. Where a measured center is compared with an absolute frequency, the free-spin line (Sec.~\ref{sm:sec:para}) and the closed form (Sec.~\ref{sm:sec:closed}), we correct it by inverting the forward model of a single line: the corrected center is the true center whose processed and fitted center equals the measured one, 3.6--3.9~GHz above it for the 7~T line (Table~\ref{sm:tab:closed}). At 8~K and 7~T the line lies at $224.6 \pm 0.5$~GHz with a FWHM of 40~GHz.

\textit{Forward model.} Every calculated line is therefore compared with the data
through the processing itself. The calculated absorption of every mode (branches,
orientations and members of the ground-state family, with their weights) is written
as a causal complex absorption $a_\pm(f)$ with intrinsic FWHM $\Gamma$. Its
field-induced change multiplies the measured zero-field trace at the same
temperature, $E(B) = E(0)\exp\{-(L/2)s[a_-(B) - a_-(0)]\}$ for the $\sigma_-$
channel and $E(B) = E(0)[t_+(B) + t_-(B)]/[t_+(0) + t_-(0)]$ with
$t_\pm = \exp(-Lsa_\pm/2)$ for the linear one, whose field is the mean of the two
circular fields; $E(B)$ and $E(0)$ are then windowed, transformed, referenced and
fitted exactly like the data, with the scale $s$ set so that the processed spectrum
peaks at 1.7~cm$^{-1}$ (3.5~cm$^{-1}$ for $\sigma_-$) within 10\% over 0.105--0.40~THz,
comparable to the measured lines. Three line shapes (a Lorentzian in the absorption,
a damped precession and a damped oscillator) and $\Gamma = 5$--20~GHz change the
estimates from the 7~T line by less than 0.4~GHz and those from the field sweeps by
up to 3.2~GHz; we use the Lorentzian with $\Gamma = 10$~GHz.

\textit{Line fitting.} Lines are fitted with a Lorentzian on a quadratic baseline over
0.105--0.40~THz. Because the mode enters the window from below, the uncertainty on the
center is set by the baseline, so every spectrum is refitted with baseline order 1, 2
and 3 and over 0.105--0.34 and 0.105--0.46~THz, and the spread is taken as the
systematic error. A center is admitted when the whole line lies in the window,
$f_0 - \Gamma/2 > 0.110$~THz and $f_0 > 0.125$~THz, the spread is below 12~GHz, and
every higher field of the same sweep is admitted too: 4--7~T at 8~K, 4.5--7~T at
9~K, 4.25--7~T at 10.5~K and 4.5--7~T in the 10~K $\sigma_-$ channel, 42 centers in
all. The last condition removes free fits at 1--3.5~T that pass the others while the
line lies below the window [grey in Fig.~3(a)]: there the free fit cannot locate the
center, and because $\Delta\alpha$ subtracts the zero-field absorption, whose tail
lies in the window, it returns centers distorted by that tail.

\subsection{Circular channels}\label{sm:sec:circular}
The incident radiation is linearly polarized along $x$ and travels along
$\bm{k}\parallel\bm{B}$. After the sample the field passes a wire-grid analyzer at
$+45^\circ$ ($D$) or $-45^\circ$ ($A$) in front of the $x$-oriented detector
antenna; since the antenna detects the $x$ projection, the two settings record
$(E_x + E_y)/2$ and $(E_x - E_y)/2$, respectively. They were measured in two runs at
10~K, from 0 to 7~T in 0.5~T steps. Each spectrum is divided by the zero-field
spectrum of its own run, $r_{D,A}(B) = S_{D,A}(B)/S_{D,A}(0)$, and
\begin{equation}
E_x = \tfrac{1}{2}\,(r_D + r_A),\qquad E_y = \tfrac{1}{2}\,(r_D - r_A),
\label{sm:eq:ExEy}
\end{equation}
in units of the zero-field $x$ field, for which $E_y(0) = 0$. The normalization
removes a difference of 12\% at 0.15--0.25~THz between the zero-field amplitudes of
the two analyzer positions and any other field-independent difference between the
runs; a zero-field dichroism is therefore not measured. The circular fields are
$E_\pm = E_x \mp iE_y$, so both helicities come from the same two measurements, and
$-(2/L)\ln|E_x|$ is the $x$-polarized absorption of the same runs. The channel that
carries the line is called $\sigma_-$ in the main text, and its identification with the Larmor sense rests on the sign of the gyromagnetic ratio.

\subsection{Dichroism estimators}\label{sm:sec:estimators}
At 10~K and 7~T the $\sigma_-$ channel peaks at 4.79~cm$^{-1}$ at 216.3~GHz and the
$\sigma_+$ channel carries $-0.05$~cm$^{-1}$ at the same frequency; the dichroism at the line center grows from 1.88~cm$^{-1}$ at 4.5~T to 4.83~cm$^{-1}$ at 7~T. At the fitted $\sigma_-$ center the dark-to-active ratio $\sigma_+/\sigma_-$ is $-0.15$ at 4.5~T and $-0.04$, $-0.02$, $+0.03$, $-0.00$ and $-0.01$ at 5--7~T, with noise 0.003--0.005.

\textit{Reference.} The circular runs absorb more than the linear sweeps: at the
$\sigma_-$ center the $x$-polarized absorption of the same runs is 1.11--1.26 times
the peak of the linear absorption interpolated to 10~K (1/3 of the 9~K and 2/3 of
the 10.5~K sweep) over 4.5--7~T. The strength of the line differs between runs by
more than its temperature dependence: at 7~T the low-temperature sweep absorbs
19--26\% less than the field sweeps at 8, 9 and 10.5~K, while within it the line
weakens by 10\% between 8 and 10.5~K. We therefore refer the channels to the $x$
channel of the same runs. The half-sum $(\sigma_- + \sigma_+)/2x$ is 1.04--1.09 over
4.5--7~T and the half-difference $(\sigma_- - \sigma_+)/2x$ is 1.41 at 4.5~T and
0.98--1.13 at 5--7~T.

\textit{Predictions.} For the two ideal cases we inject a causal Lorentzian
(intrinsic FWHM 10~GHz) at the measured $\sigma_-$ center into the zero-field traces
of the two runs, with circular weights 2 and 0 (precession about $\bm{B}$) or 1.75 and
0.25 (precession about each grain's axis, Sec.~\ref{sm:sec:powder}), scaled so that the
$x$ channel reproduces the measured $x$ absorption at the line center, and process and
fit it like the data. Over 4.5--7~T, precession about $\bm{B}$ gives a half-sum and a
half-difference of 1.02--1.08 (1.03--1.08 over 5--7~T), identical, and no dark channel;
precession about each grain's axis gives 1.01--1.04 and 0.76--0.79 (0.77--0.79 over
5--7~T) and a dark-to-active ratio of 0.14. The half-sum exceeds 1 because the $x$
channel, which transmits the mean of the two circular fields, saturates first. An
intrinsic FWHM of 20~GHz changes these values by less than 0.002.

\textit{Line-shape estimators.} The ratios above use the dark channel at one
frequency. Fitting it instead with the Lorentzian of the active line, at fixed center
and width, on a linear to cubic baseline over windows from $\pm40$~GHz about the line
to the full 0.105--0.40~THz, or with the measured active channel as template, gives
amplitudes between $-0.17$ and $+0.13$ of the active one at single fields over
5.5--7~T and between $-0.06$ and $+0.06$ averaged over these fields ($-0.09$ to $+0.01$
over 5--7~T), against 0.14 for grain axes. The spread comes from the structure of the
dark channel below the line, and we take it as the systematic uncertainty of the dark
fraction, well above the noise. If part of the absorption precessed about the grain
axes and the rest about $\bm{B}$, the line-center values would limit that part to
about a quarter, and the averaged line-shape estimators to about 45\%.

\textit{The dark channel below the line.} The dark channel carries a structure at the
lower edge of the window that does not follow the line: positive at 0.105~THz from
1.5~T on and negative near 0.15~THz at most fields, with minima of $-0.12$ to
$-0.25$~cm$^{-1}$ from 2.5~T on. At 4.5~T the active line lies on its negative part:
the dark channel at the line center is $-0.25$~cm$^{-1}$, $-0.15$ of the active one.

\textit{Single runs.} Each run alone also measures an absorption,
$-(2/L)\ln|r_{D,A}|$. The two diagonals see the same absorptive response of the sample
with opposite admixtures of the dispersive one, through the Faraday rotation that
accompanies the dichroism, so an ideal pair of runs shows mirror-image lines of nearly
equal size, shifted in opposite directions. Passed through the processing, precession
about $\bm{B}$ gives at 6--7~T a peak ratio of 1.03--1.06 between the two lines and a
separation of 28.5--31.3~GHz, precession about each grain's axis 1.02--1.04 and
22.7--24.4~GHz. The two runs give peak ratios of 0.91--1.11 and separations of
$25.9 \pm 4.4$, $29.7 \pm 3.0$ and $32.7 \pm 2.8$~GHz at 6, 6.5 and 7~T, with
$\chi^2 = 0.6$ and 13.4 for the two cases (3 points). The separation depends on the
phase of the response, not on the dark channel at the line.

\textit{Analyzer angle.} An analyzer pair at $\pm(45^\circ + \epsilon)$ scales the
reconstructed $E_y$ by $c \simeq 1 + 2\epsilon$ and lowers $D + A$ by $2\epsilon$. At zero field $D + A$ of the two runs equals the $x$-polarized transmission measured without analyzer within 1.5\% at 0.15--0.25~THz, so $|c - 1| \lesssim 0.015$, whereas grain locking would need $c = 1.33$ to produce the measured channels.

\textit{Control.}
Above $\sim\!0.5$~THz the model has no zone-center magnon. Over 0.56--1.05~THz at 8~K
and 7~T the field dependence averages 0.12~cm$^{-1}$ and peaks at 0.21~cm$^{-1}$,
against a magnon peak of 1.92~cm$^{-1}$; it grows roughly linearly with field, carries
no resonance and is absorbed by the fit baseline. 
The two circular runs are separate acquisitions, so drift between them would appear as a false dichroism; drift cannot disperse, whereas the measured dichroism follows the mode from 0.15 to 0.22~THz. 

\subsection{Paramagnetic line}\label{sm:sec:para}
Above $T_N$ the resonance continues without a break. The low-temperature sweep puts it
at 219.2~GHz at 10.5~K, just below $T_N$, and at 217.8~GHz at 11~K and
214.7--216.7~GHz at 11.5--15~K, and the temperature sweep at 211.5, 210.2, 209.8,
209.0 and 207.4~GHz at 20, 30, 40, 50 and 60~K, with an integrated intensity falling
roughly as $1/T$; above 60~K no line is resolved. Corrected for the fit bias
(Sec.~\ref{sm:sec:processing}), by inverting the forward model of a single Lorentzian injected into the zero-field trace
at the same temperature (intrinsic FWHM 15 or 25~GHz, which changes the result by
0.3~GHz), the line lies at 218.4--221.8~GHz at 11--15~K and 213.5--215.4~GHz at
20--40~K: $g_{\rm eff} = hf/\mu_B B = 2.26$ at 11~K falling to 2.18 at 40~K, 3--7\%
above the free-spin frequency of the EPR tensor, 207.0~GHz. The field of the magnet is calibrated to a few mT so the offset is not a field error. Just above $T_N$ the line thus has the $\langle g\rangle$ of the ordered-phase fits, 2.26 (Sec.~\ref{sm:sec:degeneracy}). On cooling the line moves up by $21.1 \pm 1.5$~GHz between 20 and 2.67~K.

The runs agree in temperature: the 7~T line at 10.01~K in the low-temperature sweep,
$220.1 \pm 0.8$~GHz, equals the 9.9~K point of the temperature sweep,
$220.1 \pm 0.8$~GHz, and the 8~K field sweep at 7~T, $224.6 \pm 0.5$~GHz, lies 1.0~GHz
below the 8.0~K point of the low-temperature sweep; near $T_N$ the line moves by
2--3~GHz/K, so the linear runs agree in temperature to about 0.5~K.

\subsection{Field sweeps against the model}\label{sm:sec:sweeps}
Against the calculation fitted to the 7~T line ($a = 0$, $\Delta_0 = 91$~GHz,
$\langle g\rangle = 2.26$) at the temperature of each sweep, the residuals of the
linear sweeps at 7~T are $-0.6$ to $+0.4$~GHz ($-4.0$~GHz for the $\sigma_-$ channel of
the circular runs, whose line lies 3.7~GHz below the 10.0~K linear point); toward low
field they grow to $+7$ to $+15$~GHz at 4--4.5~T [Fig.~3(a)]. The sweeps need a flatter
dispersion than the calculation. Fitted alone with one $g$ per sweep they run to the
edge of the grid for $a = 0$ and give 123 and 113~GHz for the family average and the
Y/V states (Table~\ref{sm:tab:degeneracy}). The misfit does not come from the line shape, since the three shapes and $\Gamma = 5$--20~GHz change these estimates by at most 3.2~GHz, nor from the resolution and the fit bias, which the forward model contains. The sweeps lie at 0.75--0.98~$T_N$, where the mean-field scaling with a field-independent moment is least reliable (Sec.~\ref{sm:sec:degeneracy}). The closed form of Sec.~\ref{sm:sec:closed} locates the problem: what disagrees with the calculation is the intercept, the gap at the temperature of the sweep, which falls far less toward $T_N$ than mean-field scaling
gives. Far below $T_N$, where no field sweep was measured, the two readings of the 7~T
line separate: at 2.67~K and 4~T the calculated line, processed like the data, lies at
148.2, 144.8 and 141.9~GHz for the S$^+$ fits ($a = 0$, family average, Y/V states)
and at 164.8, 159.2 and 155.5~GHz with the EPR $g$ and the gaps that reproduce the
2.67~K line, 14--17~GHz apart for each ground state, while at 7~T the two readings
agree within 0.6~GHz.

\section{Closed-form reading of the field sweeps}\label{sm:sec:closed}

Averaged over the pellet, the rigid-triad result for $\bm{B}\perp c$ extends to a
closed form for the line position,
\begin{equation}
f_0^2 = \left(\frac{\langle g\rangle\mu_B B}{h}\right)^2 + m(T)^2\Delta_0^2 ,
\label{sm:eq:closed}
\end{equation}
which reproduces the full calculation for $a = 0$ to 0.7, 1.0 and 1.5~GHz for
$\Delta_0 = 66$, 80 and 101~GHz over 4.5--7~T at the temperatures of the field sweeps.
At fixed temperature $f_0^2$ is linear in $B^2$, with slope
$(\langle g\rangle\mu_B/h)^2$ and intercept $\Delta(T)^2 = m(T)^2\Delta_0^2$. The closed form refers to the true line position, so the fitted centers are first corrected for the fit bias of the window (Sec.~\ref{sm:sec:processing}) by inverting the forward model of a single line. Straight lines through the admitted, corrected centers of each field sweep (Table~\ref{sm:tab:closedsweeps}) give $\langle g\rangle$ within 3\% of the EPR value, 2.113, and effective gaps of 68--95~GHz that fall only by a factor 1.4 between 8 and 10.5~K, where $m(T)$ falls by 1.9. With mean-field scaling they would require $\Delta_0 = 134$--185~GHz, rising toward $T_N$. The bias correction matters here, because the intercept is an extrapolation: without it the effective gaps are 6--9~GHz lower, and they still fall only by a factor 1.5. The closed form therefore reads the field sweeps as consistent with the EPR $g$ and inconsistent with the scaling; it has no ground-state degeneracy and cannot decide between the choices of
Sec.~\ref{sm:sec:degeneracy}.

\begin{table}[h]
\caption{\label{sm:tab:closedsweeps}Straight-line fits of $f_0^2$ against $B^2$ to the
admitted, bias-corrected centers of each field sweep: $\langle g\rangle$ from the
slope, the effective gap $\Delta(T)$ from the intercept, and the zero-temperature gap
that mean-field scaling would require.}
\centering
\begin{tabular}{lcccccc}
\hline\hline
Sweep & fields (T) & $\langle g\rangle$ & $\Delta(T)$ (GHz) & $m(T)$ & $\Delta(T)/m(T)$ (GHz) & $\chi^2/\nu$\\
\hline
8~K & 4--7 & $2.125 \pm 0.011$ & $95 \pm 2$ & 0.709 & $134 \pm 2$ & 1.2\\
9~K & 4.5--7 & $2.153 \pm 0.013$ & $85 \pm 2$ & 0.631 & $135 \pm 3$ & 1.4\\
10~K, $\sigma_-$ & 4.5--7 & $2.092 \pm 0.016$ & $81 \pm 3$ & 0.506 & $160 \pm 6$ & 3.9\\
10.5~K & 4.25--7 & $2.168 \pm 0.017$ & $68 \pm 4$ & 0.370 & $185 \pm 11$ & 1.4\\
\hline\hline
\end{tabular}
\end{table}

\end{document}